\documentclass[conference]{IEEEtran}
\IEEEoverridecommandlockouts
\usepackage{cite}
\usepackage{amsmath,amssymb,amsfonts}
\usepackage{algorithmic}
\usepackage{graphicx}
\usepackage{textcomp}
\usepackage{xcolor}
\usepackage{array}
\usepackage{multirow}
\usepackage{amssymb}
\usepackage{amsmath}
\usepackage{booktabs}
\usepackage{ulem}
\usepackage{verbatim}
\usepackage{url}
\usepackage{color}
\usepackage{verbatim}
\usepackage[ruled,linesnumbered]{algorithm2e}
\def\BibTeX{{\rm B\kern-.05em{\sc i\kern-.025em b}\kern-.08em
    T\kern-.1667em\lower.7ex\hbox{E}\kern-.125emX}}
    
\makeatletter
\newcommand{\linebreakand}{%
  \end{@IEEEauthorhalign}
  \hfill\mbox{}\par
  \mbox{}\hfill\begin{@IEEEauthorhalign}
}
\makeatother
\begin{document}

\title{Multi-modal Dynamic Graph Network: Coupling Structural and Functional Connectome for Disease Diagnosis and Classification \\
\thanks{* Corresponding authors.}
}

\author{

\IEEEauthorblockN{1\textsuperscript{st}Yanwu Yang}
\IEEEauthorblockA{
\textit{Harbin Institute of Technology} \\ \textit{at Shenzhen }\\
Shenzhen, China \\
\textit{Peng Cheng Laboratory}\\
Shenzhen, China}
\and

\IEEEauthorblockN{2\textsuperscript{rd}Xutao Guo}
\IEEEauthorblockA{
\textit{Harbin Institute of Technology} \\ \textit{at Shenzhen }\\
Shenzhen, China \\
\textit{Peng Cheng Laboratory}\\
Shenzhen, China}
\and

\IEEEauthorblockN{3\textsuperscript{th} Zhikai Chang}
\IEEEauthorblockA{
\textit{Harbin Institute of Technology} \\ \textit{at Shenzhen }\\
Shenzhen, China}
\and

\linebreakand

\IEEEauthorblockN{4\textsuperscript{th} Chenfei Ye}
\IEEEauthorblockA{
\textit{Harbin Institute of Technology} \\ \textit{at Shenzhen }\\
Shenzhen, China}
\and

\IEEEauthorblockN{5\textsuperscript{th} Yang Xiang*}
\IEEEauthorblockA{\textit{Peng Cheng Laboratory} \\
Shenzhen, China}
\and

\IEEEauthorblockN{6\textsuperscript{th} Ting Ma*}
\IEEEauthorblockA{
\textit{Harbin Institute of Technology} \\ \textit{at Shenzhen }\\
Shenzhen, China \\
\textit{Peng Cheng Laboratory}\\
Shenzhen, China}
}

\maketitle

\begin{abstract}
Multi-modal neuroimaging technology has greatlly facilitated the efficiency and diagnosis accuracy, which provides complementary information in discovering objective disease biomarkers. Conventional deep learning methods, e.g. convolutional neural networks, overlook relationships between nodes and fail to capture topological properties in graphs. Graph neural networks have been proven to be of great importance in modeling brain connectome networks and relating disease-specific patterns. However, most existing graph methods explicitly require known graph structures, which are not available in the sophisticated brain system. Especially in heterogeneous multi-modal brain networks, there exists a great challenge to model interactions among brain regions in consideration of inter-modal dependencies. In this study, we propose a Multi-modal Dynamic Graph Convolution Network (MDGCN) for structural and functional brain network learning. Our method benefits from modeling inter-modal representations and relating attentive multi-model associations into dynamic graphs with a compositional correspondence matrix. Moreover, a bilateral graph convolution layer is proposed to aggregate multi-modal representations in terms of multi-modal associations. Extensive experiments on three datasets demonstrate the superiority of our proposed method in terms of disease classification, with the accuracy of 90.4\%, 85.9\% and 98.3\% in predicting Mild Cognitive Impairment (MCI), Parkinson's disease (PD), and schizophrenia (SCHZ) respectively. Furthermore, our statistical evaluations on the correspondence matrix exhibit a high correspondence with previous evidence of biomarkers.


\end{abstract}


\section{Introduction}
Recently, computer-aided diagnosis technologies using advanced neuroimaging developments have been widely adopted for medical scenarios, e.g. disease diagnosis, and medical image segmentation.
Among these neuroimaging tools, functional Magnetic Resonance Imaging (fMRI) and Diffusion Tensor Imaging (DTI) have become promising candidates for brain study.
Functional MRI is a stimulus-free acquisition used to track changes in co-activation across brain regions. DTI captures the directional diffusion of water molecules as a proxy for structural connectivity.
Derived functional and structural connectivity is feasible to model the brain as a network by representing brain parcellations along with their structural or functional connectivity. The brain connectome provides a more holistic view by modeling the entire human brain and characterizes individual subject behavior, cognition, and mental health \cite{dadi2019benchmarking}.
There is mounting evidence that demonstrates functional and structural connectivity could be used to identify predictive biomarkers for brain disorders such as Alzheimer's disease (AD), Schizophrenia (SCZ), and Parkinson's disease (PD) \cite{yang2021alteration,ye2019connectome,damaraju2014dynamic}.

Medical image-based diagnosis is a challenging task due to the sophisticated structure of brain systems and subtle lesions, which might be overlooked by medical experts \cite{yang2020deep}. Neuroimage processing with multiple modalities is feasible to assess and develop distinctive biomarkers from multiple fields.
Previous studies link the functional signals with structural pathways for mediating and suggest that functional connectivity and structural connectivity might be mediated by each other \cite{fukushima2018structure,atasoy2016human,dsouza2021m}.
Recently, state-of-the-art graph neural networks (GNN) have achieved promising performance in multi-modal graph-structural data learning \cite{dsouza2021m,jiang2019dynamic,saqur2020multimodal}.
However, most existing GNNs built graphs on the originally derived connectivity, and fail to sufficiently model sophisticated associations among nodes.
This issue is even aggravated when modeling multi-modal brain networks since there exist heterogeneous structures and representations among multiple modalities. However, most existing studies potentially ignore these issues and achieves sub-optimal results.

 


In terms of this, we propose a Multi-modal Dynamic Graph Convolution Network (MDGCN) to model multi-modal complementary associations by dynamic graphs. Our network allows for tighter coupling of context between multiple modalities by representing functional and structural connectome dynamically and providing a compositional space for reasoning. Specially, we first parse both the functional and structural into dynamic graphs with embedded representations as nodes. A correspondence factor matrix is introduced to capture the corresponding values of each pair of nodes between modalities, which is denoted as the adjacency matrix. And multimodal representations are aggregated by a Bilateral Graph Convolution (BGC) layer for complementary message passing. Extensive experiments on three datasets demonstrate that our proposed method outperforms other baselines in the prediction of Mild Cognitive Impairment (MCI), Parkinson's Disease (PD), and Schizophrenia (SCHZ) with the accuracy performances of 90.4\%, 85.9\%, and 98.3\% respectively.

The rest of our paper is structured as follows. We would like to review competitive methods in terms of connectome study and multi-modal models in Section \ref{related works}. The details of the proposed model are introduced in Section \ref{method}. Section \ref{res} describes the experiments of our proposed model in disease classification on 3 datasets and provides the experimental results. Section \ref{con} draws the conclusions of the work.

\section{Related works}\label{related works}
\subsection{Brain connectome network study}
With the flexibility of uncovering the complex biological mechanisms using rs-fMRI and DTI, deep learning methods have been widely coordinated to examine and analyze the patterns. Convolution neural networks (CNN) and graph neural networks (GNN) have become useful tools for brain connectome embedding, where high dimensional neuroimaging features are embedded into a low dimensional space that preserves their context as well as capturing topological attributes. BrainNetCNN is proposed to take the brain connectome networks as grid-like data, and measure the topological locality in connectome \cite{kawahara2017brainnetcnn}, which has achieved promising performance for disease diagnosis and phenotype prediction.

Apart from convolution neural networks, graph neural networks retain a state that can represent information about the neighbors and provides a powerful way to explore the dependencies between nodes. However, applying a graph network directly to the brain connectome is problematic. On one hand, brain networks have sophisticated and non-linear structures.
For example, most existing methods apply the derived functional connectivity as the adjacency matrix, which is measured linearly between two brain regions. These derived linear connectivities fail to model complex associations between brain regions. On the other hand, graph convolution networks explicitly require a known graph structure, which is not available in the brain connectome. Several strategies have been proposed to tackle the unknown structure issue \cite{hu2020class,zhang2019new,lei2020self}. Especially, dynamic graph convolution methods are proposed to model graph structures adaptively to characterize intrinsic brain connectome representations and achieve promising performances in prediction \cite{zhao2022dynamic, jiang2019dynamic}. Nevertheless, there is still a lack of studies to tackle the multi-modal connectome graphs.

\subsection{Multi-modal connectome learning}
Existing multi-modal connectome learning methods can be categorized into two classes: feature learning methods and deep learning methods. Compared with feature learning methods \cite{li2020deep,chu2012does,salas2010feature} that leverage feature selection to identify disease-related features, deep learning methods are feasible to capture intrinsic meaningful representations and achieves better performances. \cite{song2021graph} devised a calibration mechanism to fuse fMRI and DTI information into edges. \cite{wang2018novel} proposed to perform a two-layer convolution on the fMRI and DTI data simultaneously. \cite{dsouza2021m} regularizes convolution on functional connectivity with structural graph Laplacian. However, most of these studies lack the ability to sufficiently model complementary associations between modalities, since there is a lack of joint compositional reasoning over both functional and structural connectome networks.

\section{Method}\label{method}
The proposed Multi-modal Dynamic Graph Convolution Network (MDGCN) aims at parsing multi-modal representations into dynamic graphs and performing graph aggregation for message passing. In this section, we would like to firstly introduce the brain graph definition, and then the detail of the proposed method.

\subsection{Preliminaries}
\textbf{Brain network graph: }
The brain networks derived from neuroimages are usually symmetric positive define (SPD) matrices $X \in \mathbb{R}^{M \times M}$, where $M$ denotes the number of brain regions. Each element $x_{i,j}$ denotes a co-variance or connectivity strength between the regions. 
The brain network is usually formulated as an undirected graph $G=(V,E,H)$, where $V$ is a finite set of vertices with $|V|=M$ and $E \in \mathbb{R}^{M \times M}$ denotes the edges in the graphs. The nodes and edges are represented by the derived SPD matrices $X$. For each vertex $v_i$, the node feature vector $h_i$ is constructed by the $i$-th row or column in the SPD matrix $h_i = \{x_{i,k}|k=1,2,...,M\}$. The edges are represented by the matrices directly, of which an element is assigned by $e_{i,j}=x_{i,j}$.

\begin{figure*}
    \centering
    \includegraphics[scale=0.95]{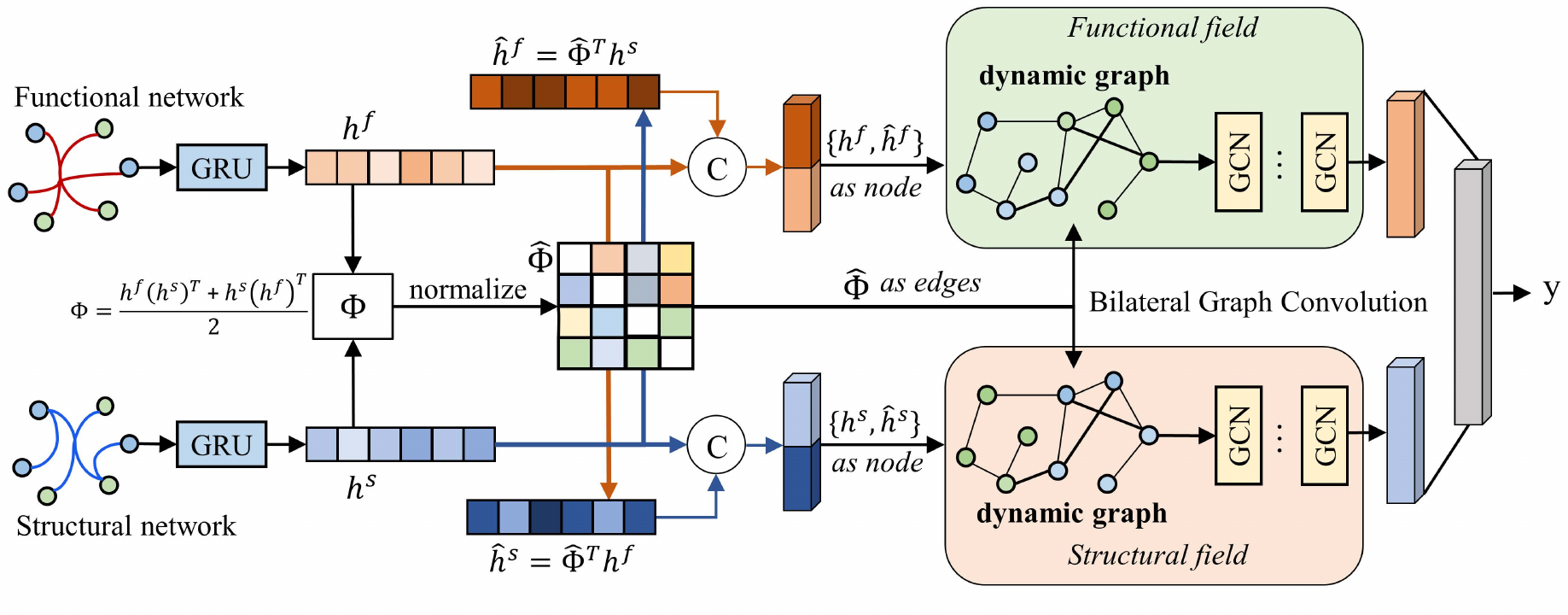}
    \caption{Illustration of the proposed MDGCN method, where GRU denotes a gated recurrent unit for node representation learning, GCN denotes a graph convolution layer. The multi-modal representations are cross embedded and fed into a bilateral graph convolution module. Message passing on the graphs are guided by the correspondence matrix.}
    \label{figureconv}
\end{figure*}

\textbf{Multi-modal brain graph: }
The multi-modal brain graphs are constructed by the functional and structural brain networks derived from fMRI and DTI respectively. An input $\hat{G}$ is expressed by a tuple of graphs as $\hat{G}=\{G^s,G^f\}$, where $G^s$ and $G^f$ denote the structural and functional brain network graphs respectively. Formally, given a set of graphs $\{\hat{G}_1,\hat{G}_2,...,\hat{G}_N\}$ with a few labeled graph instances, the aim of the study is to decide the state of the unlabeled graphs as a graph classification task.


\textbf{Dynamic graph: }
Since there remain unknown node relationships in the multi-modal graphs, applying the graph convolution on the multi-modal graphs is problematic. 
The dynamic graphs leverage the dynamic mechanism to model multi-modal node representations and interactions by learning mappings: $f_V:X\rightarrow V, f_E:X \rightarrow E$. And the learned graphs are denoted as dynamic graphs.

\subsection{Dynamic Multi-modal Graph Mapping}
The key to brain graph mapping is how to parse multi-modal graph representations. In this study, we propose to model the node representations as a sequence via the Gated Recurrent Unit (GRU) \cite{chung2014empirical}, where the gating mechanism allows for the learning of sequential relationships and protect the learning from undesired updates \cite{hochreiter1997long}. 
The GRU layers take the input brain network matrix $X \in \mathbb{R}^{M \times M}$ as a sequence with $M$ nodes and $M$ features.
Formally, given an input $X = \{X^f,X^s\}$, where $X^f$ and $X^s$ denotes the functional and structural brain networks, the node features are embedded by:
\begin{equation}
h^s_j = GRU(\hat{x}^s), h^f_j = GRU(\hat{x}^f)
\end{equation}
Specially, given an input $\hat{x}$, the $GRU$ operation on each parcel $k \in [1,M]$ is formulated as:
\begin{equation}
z_k = \sigma(W_z \cdot[o_{k-1},\hat{x}_k])
\end{equation}

\begin{equation}
r_k = \sigma(W_r \cdot[o_{k-1},\hat{x}_k])
\end{equation}

\begin{equation}
\hat{h}_k = tanh(W \cdot[r_k * o_{k-1},\hat{x}_k])
\end{equation}

\begin{equation}
o_k = (1-z_k)*o_{k-1}+z_k*\hat{o}_k
\end{equation}

where, $\hat{x}_k$ denotes the input vector corresponding to the $k$-th brain region, and $z_k$, $r_k$ are the update gate and reset gate respectively.
With the embedded $h^f_j$ and $h^s_j$, a soft correspondence matrix $\Phi$ is obtained by:
\begin{equation}\label{phi}
\Phi = \frac{h^f_j (h^s_j)^T + h^s_j (h^f_j)^T}{2}
\end{equation}

Each row vector in $\Phi$ is a probability distribution over potential correspondences to corresponding nodes. The matrix can be regarded as the scores for measuring the goodness of matches between nodes in two modalities.
A sinkhorn function is applied to normalize the matrix, which satisfies doubly stochastic, where $\sum_{j}^{M} \hat{\Phi}_{i,j} =1$.

By obtaining the normalized correspondence matrix $\hat{\Phi}$, we can project the representations from one source into another source by:
\begin{equation}
\hat{h}^f = \hat{\Phi}^T h^s
\end{equation}
\begin{equation}
\hat{h}^s = \hat{\Phi}^T h^f
\end{equation}

With the obtained embedded representations $\hat{h}^f$ and $\hat{h}^s$, the dynamic node features can be built by $\{\hat{h}^f,h^f\}$ or $\{\hat{h}^s,h^s\}$, which represents the transformation representations of functional/structural field. Moreover, we formulate the normalized correspondence matrix $\hat{\Phi}$ as the dynamic adjacency matrix.

\subsection{Bilateral Graph Convolution}
In this study, a Bilateral Graph Convolution (BGC) is proposed to perform convolution on the multi-modal graphs. To tackle the heterogeneous features between modalities, the BGC module applies convolutions on each field to aggregate representations of each single modality separately. Moreover, we implement the spatial aggregation on the graphs for message passing instead of spectral graph convolution. Since the brain network is fully connected, graph spatial convolution as well as spectral graph convolutions are able to aggregate global information. In this way, the graph spatial convolution is formulated with $\hat{\Phi}$ as:

\begin{equation}
H^f_{l+1} = \sigma(\hat{\Phi}H^f_lW^f_l), \text{where } H^f_0=||\{\hat{h}^f,h^f\}
\end{equation}
\begin{equation}
H^s_{l+1} = \sigma(\hat{\Phi}H^s_lW^s_l), \text{where } H^s_0=||\{\hat{h}^s,h^s\}
\end{equation}
$||$ denotes a concatenation operation, $\sigma$ denotes a sigmoid activation function, and $W$ is a learnable matrix for improving node representations. The outputs of the BGC layer are further combined as a feature vector, and fed into a multi-layer perception for classification.

\subsection{Optimization}
The final output is further fed into a three-layer multi-layer perception classifier followed with a ReLU activation function and a dropout layer. A softmax function is implemented for probability output. The loss function is formulated as the cross-entropy function. To summarize, the detail of the process of our proposed DMGN is shown in Algorithm \ref{alg}:
\begin{algorithm}\label{alg}
\caption{Dynamic Multi-modal Graph Network}
\LinesNumbered
\KwIn{Multi-modal brain networks $\{\hat{G}_{1}, \hat{G}_{2}, \hat{G}_{3}, ..., \hat{G}_{n}\}$,where $\hat{G}=\{G^s,G^f\}$; For each graph $G=(V,E,X)$}
\KwOut{Prediction $y$ of the test set}
Calculate dynamic node features by GRU:
$h^s = GRU(\hat{x}^s)$, $h^f = GRU(\hat{x}^f)$\;
Calculate the correspondence matrix: $\Phi = \frac{1}{2}h^f_j (h^s_j)^T + \frac{1}{2}h^s_j (h^f_j)^T$\;
Normalize $\Phi$ into $\hat{\Phi}$\ to satisfy doubly stochastic\;
Obtain the cross-modality mapping representations: \\ $\hat{h}^f = \hat{\Phi}^T h^s$\; $\hat{h}^s = \hat{\Phi}^T h^f$\;

\For{l=1:L}{	
$H_{l}^f = \sigma(\hat{\Phi}H_{l-1}^fW^f_{l-1})$, where $H^f_0 = ||\{\hat{h}^f,h^f\}$\;
$H_{l}^s = \sigma(\hat{\Phi}H_{l-1}^sW^s_{l-1})$, where $H^s_0 = ||\{\hat{h}^s,h^s\}$\;
}
Readout $H_L = ||\{H_{L}^f,H_{L}^s\}$\;
$y=softmax(H_L)$
\end{algorithm}

\section{Experiments and Results}\label{res}
\subsection{Datasets}
In this study, three real-world datasets are employed in this study, where functional MRI and DTI are aggregated. All the datasets are enrolled for multi-modal graph classification.

\textbf{ADNI Dataset}\footnote{\url{http://www.adni-info.org/}}: The ADNI dataset is a longitudinal multimodal neuroimaging dataset. In this study, we collected 114 subjects that were diagnosed at the baseline for evaluation including 51 healthy controls (NC) and 63 mild cognitive impairment (MCI). Notably, MCI is considered to be a significant stage for preclinical diagnosis of AD.

\textbf{Xuanwu dataset} \cite{yang2021alteration}:  A total of 155 subjects are included in this dataset, where 70 HCs and 85 subjects with Parkinson's Disease (PD) were recruited from the Movement Disorders Clinic of the Xuanwu Hospital of Capital Medical University. The patients were diagnosed according to the MDS Clinical Diagnostic Criteria for Parkinson's disease.

\textbf{CHUV dataset} \cite{van2013wu}: The MRI data were obtained from the Service of General Psychiatry at the Lausanne University Hospital including 27 healthy participants and 27 schizophrenic patients. All of the patients were diagnosed with schizophrenic disorders after meeting the DSM-IV criteria.
\begin{table*}[htbp]
  \centering
  \caption{Multi-modal classification results across 10-fold cross validation (mean \%$\pm$ std\%) in terms of accuracy (Acc), precision (Prec) and area under the curve (AUC). The best results for each column is shown in bold, and the second best is in underline.}
\renewcommand\arraystretch{1.3}
    \begin{tabular}{c|ccc|ccc|ccc}
    \hline
    \multirow{2}[3]{*}{Method} & \multicolumn{3}{c|}{ADNI} & \multicolumn{3}{c|}{Xuanwu} & \multicolumn{3}{c}{CHUV} \\
\cline{2-10}          & Acc   & Prec  & AUC   & Acc   & Prec  & AUC   & Acc   & Prec  & AUC \\
\hline
    SVM   & 63.2$\pm$9.5 & 60.8$\pm$6.5 & 68.3$\pm$13.6 & 62.6$\pm$3.3 &\uline{90.6$\pm$2.5} & 69.8$\pm$5.1 & 64.7$\pm$4.2 & 63.4$\pm$7.0 & 72.9$\pm$15.9 \\

   M-MLP & 80.6$\pm$9.9 & 76.29$\pm$12.3 & 83.9$\pm$9.8 & 73.8$\pm$7.4 & 70.2$\pm$6.4 & 73.6$\pm$9.7 & 82.7$\pm$11.1 & 80.0$\pm$18.9 & 82.4$\pm$11.6 \\

    BrainNetCNN & 82.3$\pm$11.5 & 82.1$\pm$14.1 & 82.8$\pm$9.5 & 75.6$\pm$9.7 & 71.4$\pm$12.1 & 77.6$\pm$10.7 & 92.7$\pm$9.0 & \uline{95.0$\pm$10.0} & 94.7$\pm$8.7 \\

    M-GCN & 83.3$\pm$7.4 & 80.9$\pm$15.9 & 81.7$\pm$13.6 & \uline{81.3$\pm$4.6} & 77.1$\pm$11.3 & 75.4$\pm$10.2 & 90.7$\pm$12.9 & 94.2$\pm$11.8 & 86.8$\pm$19.7 \\

    HGNN  & 81.1$\pm$6.5 & \uline{84.8$\pm$14.2} & 85.6$\pm$6.4 & 77.7$\pm$9.8 & 76.9$\pm$16.2 & 80.3$\pm$12.2 & \uline{96.0$\pm$8.0} & \boldmath{}\textbf{100.0$\pm$0.0}\unboldmath{} & \uline{98.9$\pm$0.9} \\

    DHGNN & \uline{84.4$\pm$6.0} & 84.2$\pm$12.3 & \uline{86.7$\pm$5.4} & 78.2$\pm$6.8 & 80.1$\pm$13.1 & \uline{80.8$\pm$8.1} & 90.7$\pm$9.4 & 83.3$\pm$17.9 & 92.4$\pm$10.8 \\

    DMGN(Ours)  & \boldmath{}\textbf{90.4$\pm$2.4}\unboldmath{} & \boldmath{}\textbf{91.6$\pm$6.9}\unboldmath{} & \boldmath{}\textbf{88.2$\pm$6.6}\unboldmath{} & \boldmath{}\textbf{85.9$\pm$4.5}\unboldmath{} & \boldmath{}\textbf{85.5$\pm$6.0}\unboldmath{} & \boldmath{}\textbf{83.1$\pm$8.1}\unboldmath{} & \boldmath{}\textbf{98.3$\pm$5.0}\unboldmath{} & \boldmath{}\textbf{100.0$\pm$0.0}\unboldmath{} & \boldmath{}\textbf{99.2$\pm$1.7}\unboldmath{} \\
    \hline
    \end{tabular}%
  \label{tab1}%
\end{table*}%

\subsection{Preprocessing}
All the fMRI images were pre-processed by reference to the Configurable Pipeline for the Analysis of Connectomes (CPAC) pipeline \cite{craddock2013towards}, including skull striping, slice timing correction, motion correction, global mean intensity normalization, nuisance signal regression with 24 motion parameters, and band-pass filtering (0.01-0.08Hz). The functional images were finally registered into a standard anatomical space (MNI152). The mean time series for a set of regions were computed and normalized into zero mean and unit variance. The Pearson Coefficient Correlation is applied to measure the functional connectivity.

The DTI images were pre-processed by image denoising, head motion, eddy-current, susceptibility distortion, and field inhomogeneity correction by MRtrix 3 \cite{tournier2012mrtrix}. The fiber orientation distributions were estimated by constrained spherical deconvolution \cite{jeurissen2014multi}. We performed the 2-nd order Integration over Fiber Orientation Distributions \cite{anderson2005measurement} to reconstruct 10 million streamlines. A Spherical-deconvolution Informed Filtering of Tractograms \cite{smith2013sift} was applied to reduce the streamline count to 5 million. The number of streamlines connecting each pair of brain regions was used to construct the structural network.

All the pre-processed fMRI and DTI images were mapped by the brain template for parcellations. In this study, the images in ADNI and Xuanwu datasets were segmented by the Schaefer atlas \cite{schaefer2018local}, which was parceled by a gradient weighted Markov random field approach that identified 100 cortical parcels. CHUV data were parcellated into 83 cortical and subcortical areas by the Freesurfer \cite{fischl2012freesurfer}.

\subsection{Implementation details}
In our implementation, the number of layers of graph convolution is decided in a grid search from 1 to 4. The output of graph convolution is further fed into a 3-layer multi-layer perception classifier followed by a leaky ReLU activation function and a dropout layer. The learning rate is set as 3e-4, and the weight decay is 5e-5. All the models in this study are trained for 600 epochs and would be early stopped then the loss has not been decreased for 100 epochs. We trained the models with PyTorch on two NVIDIA 2080-Ti GPUs. For better comparison, 10-fold cross-validation was applied for evaluation by randomly sampling 90\% data for training and 10\% for testing in each fold. For all experiments, we evaluated the performance in terms of the diagnosis accuracy (Acc), precision (Prec), and Area Under the Curve (AUC).

\subsection{Competitive methods}
In this study, we compare our proposed MDGCN with well-estimated graph methods and models that are designed specifically for brain connectome. These methods include:

\textbf{SVM and MLP.} The conventional machine learning methods of the support vector machine and multi-layer perception are compared as a baseline of classification. The upper matrix of the brain networks are fed into the classifiers to give a score for each subject. The layer number of MLP is searched from 1 to 4.

\textbf{BrainNetCNN} \cite{kawahara2017brainnetcnn}. BrainNetCNN is a CNN-based framework for brain network study with promising performances in brain network study. The model is implemented by multiple convolution layers to learn the multi-modal inputs.

\textbf{M-GCN} \cite{dsouza2021m}. M-GCN is a multi-modal graph convolution network that aggregates functional representations by regularizing with structural graph Laplacian, which outperforms several stat-of-the-art baselines in predicting phenotypic values.

\textbf{HGNN} \cite{feng2019hypergraph}. A hypergraph graph neural network is implemented to encode hyper structure. The multi-modal brain networks can be fed into the HGNN as a nature of hyperedges.

\textbf{DHGNN} \cite{jiang2019dynamic}. The dynamic graph hyperneural network extends the HGNN into a dynamic graph. These two methods are implemented to compare with the hyperedge mechanism for modeling multi-modal data.

\subsection{Parameter setting}
\begin{figure}
    \centering
    \includegraphics[scale=0.37]{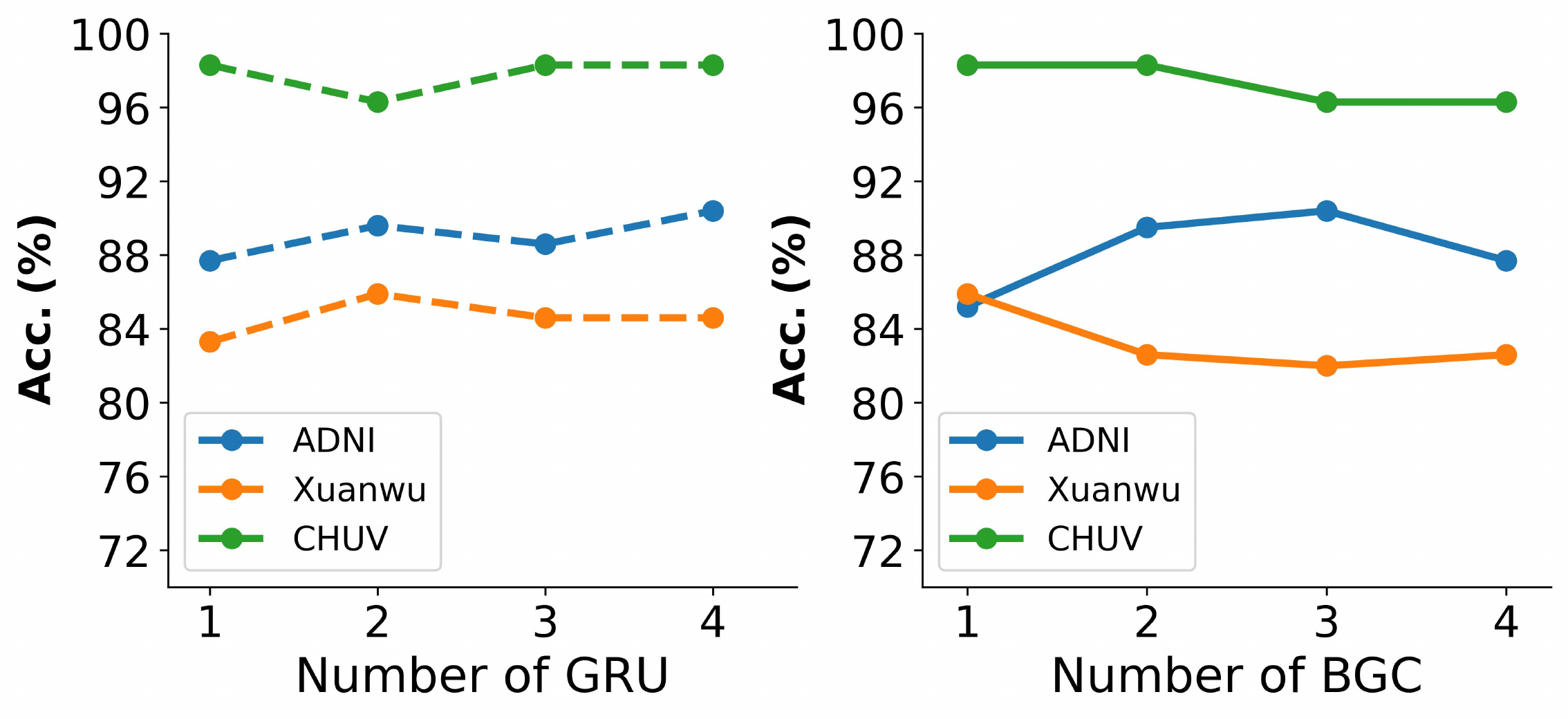}
    \caption{The effect of the number setting for the GRU layer and Bilateral Graph Convolution layer.}
    \label{param}
\end{figure}
The settings of the GRU layer and the bilateral graph convolution layer are discussed in Figure \ref{param}, where the results on the ADNI, Xuanwu and CHUV datasets are shown in blue, yellow, and green respectively. The prediction performances would slightly change with the setting on the layer number of GRU. Moreover, as the layer number of bilateral graph convolution increases, the performance trends to increase firstly and then decrease. We suspect that this is caused by the global aggregation of the graph convolution, which leads to over-smoothing on graphs.

\begin{figure}
    \centering
    \includegraphics[scale=0.4]{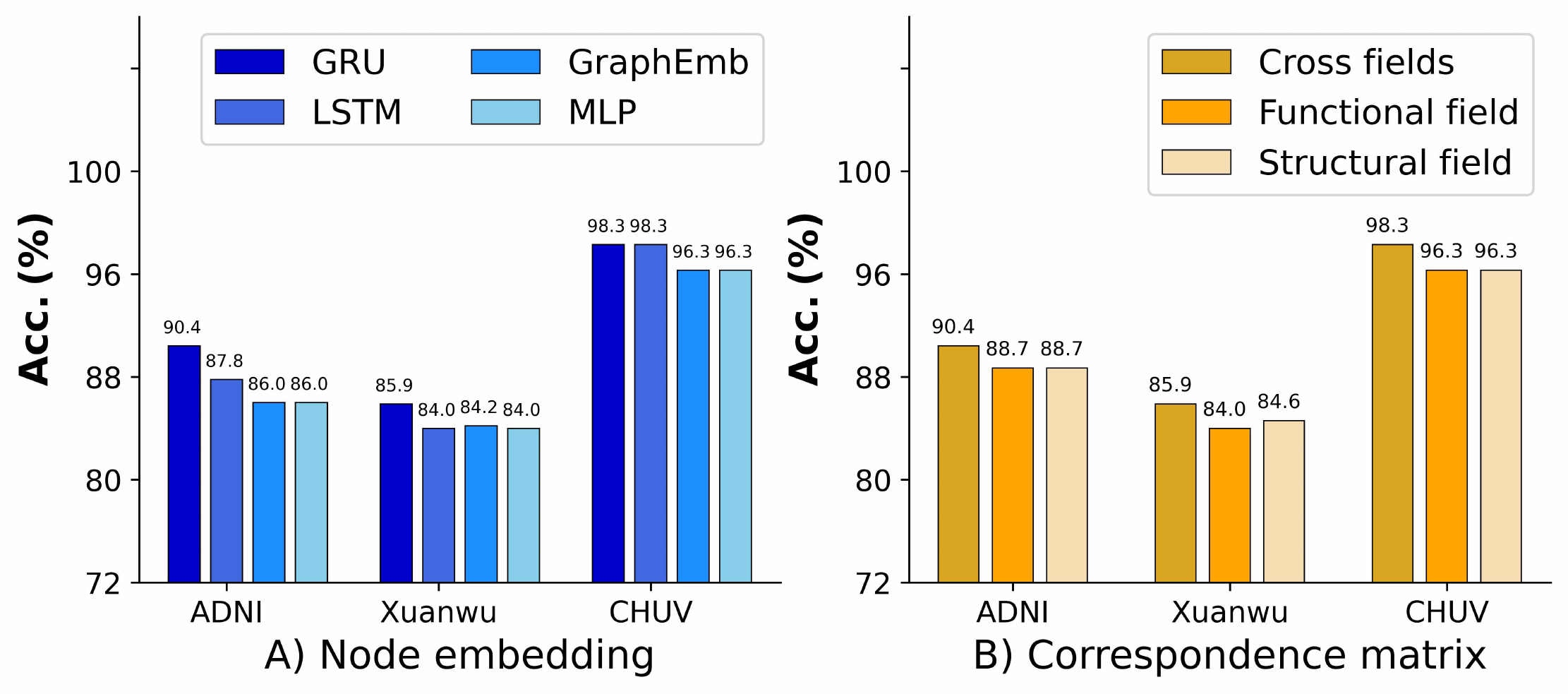}
    \caption{Comparison results in terms of the node embedding and cross graph convolution.}
    \label{ablation}
\end{figure}

\begin{figure*}
    \centering
    \includegraphics[scale=0.93]{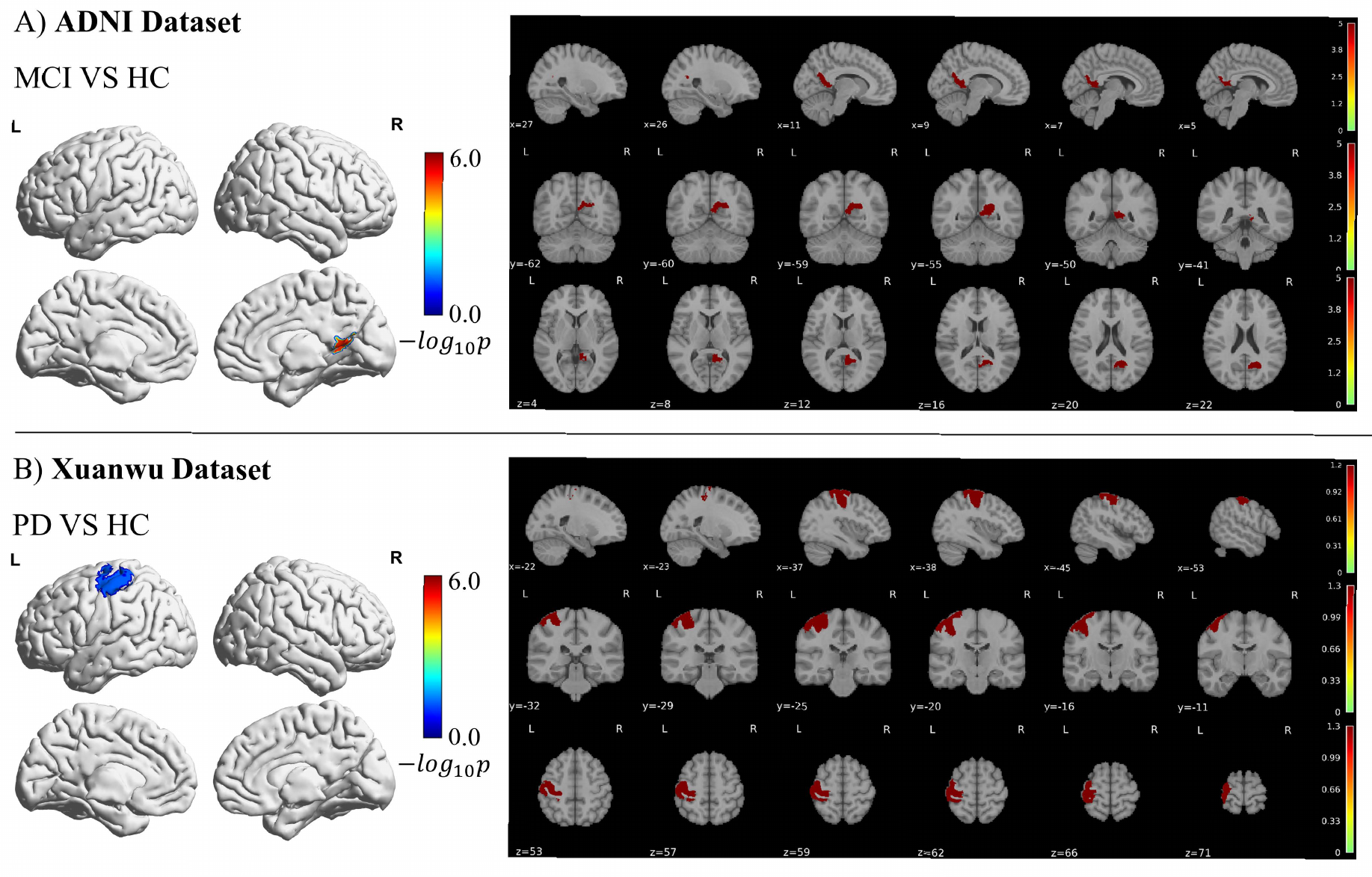}
    \caption{Statistical results ($-log_{10} p$) on the correspondence matrix, where only significant regions are displayed. The right posterior cingulate gyrus in the default mode network, the left somatomotor network are located in MCI and PD respectively.}
    \label{explain}
\end{figure*}

\subsection{Multi-modal graph classification performance}
Table \ref{tab1} demonstrates the comparison results, where the best results are shown in bold, and the second best are in underline. It is shown that conventional machine learning SVM achieves the worst performance. This indicates that the multi-modal graphs might have nonlinear and heterogeneous structures, and a simply linear classifier fails to distinguish the brain disease state. Moreover, compared with SVM, multi-layer perception improves the performances significantly, where nonlinear representations are deeply embedded with multiple layers. BrainNetCNN and M-GCN achieve further improvements, showing that deep learning methods such as CNN and GNN are feasible to capture and learn more meaningful representations. In addition, the hypergraph neural network HGNN performs worse than BrainNetCNN and M-GCN on the ADNI and Xuanwu datasets. While DHGNN outperforms these baseline models on the ADNI and Xuanwu datasets, where the dynamic mechanism might contribute to richer representations. Moreover, our proposed MDGCN achieves the best performances among all the competitive models in terms of the accuracy, precision and AUC, with 6.0\%, 4.6\% and 2.3\% improvements on the ADNI, Xuanwu and CHUV datasets respectively. We suspect that the improvements are caused by the powerful cross-modality correspondence reasoning that captures inter-modal dependencies to aggregate multi-modal representations.

\subsection{Ablation studies}

Ablation studies were carried out to measure the importance of the components in the MDGCN, including edge-aware convolution and the correspondence matrix. Figure \ref{ablation} plots the comparison results on the way of node embedding and the correspondence matrix. On one hand, the GRU for node embedding is compared with other node embedding methods, including multi-layer perception, graph embedding, and LSTM \cite{hochreiter1997long}. As is shown in Figure \ref{ablation} A), the LSTM outperforms the graph embedding and MLP methods. Moreover, the GRU performs better than LSTM on the ADNI and Xuanwu datasets. The results demonstrate that the gating mechanism in LSTM and GRU could relate long-term attentive relationships, capture more meaningful node representations, and protect the node representations from undesired updates, which plays a key role in dynamic graph node representation learning. The graph embedding and MLP embed all the node features in the same and fails to consider the importance of connectivity patterns. 
On the other hand, we also evaluated the effect of the correspondence matrix by comparing with the single domain aggregation. In detail, the equation (\ref{phi}) is modified as $\Phi = h^f_j (h^s_j)^T$ and $\Phi = h^s_j (h^f_j)^T$ to calculate the function-structure and structure-function transformation respectively. And only a single domain aggregation is read out by $H_L=H_L^f$ or $H_L=H_L^s$. 
Figure \ref{ablation} B) displays the comparison results. It is shown that results of graph aggregations in the functional domain and structural domain are comparable in terms of the accuracy. Moreover, our method incorporating cross-domain aggregation outperforms the single domain aggregation. This indicates that the interplay between modalities might be a key in modeling multi-modal representations. 

\subsection{Biological explanation}
The correspondence matrices attentively model the multi-modal interactions among regions, which are feasible to be applied to pinpoint the key brain biomarkers for disease. In this section, we extracted the correspondence matrix for each subject and performed a group statistical analysis for the explanation. The experiments were carried out on the ADNI and Xuanwu datasets that consist of more than 100 subjects for example, which could tell the power of the proposed DMGCN in interpretation ability.
In detail, we performed the state-of-the-art statistical method, multi-distance multi-variant regression (MDMR), for group comparison on the correspondence matrix. The Bonferroni correction was applied to control the false positive rate. And the p value $< 0.05$ after the correction was determined significant within the experiments.

Figure \ref {explain} demonstrates the statistical analysis of the two datasets, where the regions with significant differences (p value$ <0.05$) are displayed with the value of $-log_{10}p$. The regions of the right posterior cingulate gyrus in the default mode network and the left somatomotor were found in the ADNI, and Xuanwu datasets respectively. 
Previous studies demonstrate that the AD process has been hypothetically explained by PCC hypofunction, due to the effect of the degeneration of cingulum fibers \cite{choo2010posterior}. And PCC hypofunction could be caused by early PCC atrophy, which has been proven to be a useful biomarker \cite{chua2009diffusion}. Moreover, Parkinson's Disease is shown to be beginning to develop somatomotor dysfunctions with deficits in neocortical association areas \cite{braak2006stanley}. Our findings in the right posterior cingulate gyrus of the default mode network, and the left somatomotor network coincide with these previous studies, indicating that our proposed MDGCN could locate meaningful and interpretative key biomarkers.

\section{Conclusion}\label{con}
In this study, we propose a multi-modal dynamic graph convolution network for disease diagnosis and classification, which exploits a dynamic graph for tighter coupling of multi-modal representations. The derived correspondence matrix provides a compositional space for reasoning multi-modal dependencies. The experimental results on three datasets demonstrate that our proposed method is feasible to model multi-modal graphs and outperforms other state-of-the-art methods. Moreover, by performing statistical analysis on the correspondence matrix, the high correspondence with previous evidence of biomarkers exhibits the interpretation ability of our proposed MDGCN, which provides a powerful way of multi-modal brain network learning.

\section*{Acknowledgment}
This study is supported by grants from the the National Natural Science Foundation of China (62106113, 6227608 and 62106115), the Innovation Team and Talents Cultivation Program of National Administration of Traditional Chinese Medicine (NO:ZYYCXTD-C-202004), Basic Research Foundation of Shenzhen Science and Technology Stable Support Program (GXWD20201230155427003-20200822115709001), The National Key Research and Development Program of China (2021YFC2501202), and the Major Key Project of PCL.

\normalem
\bibliographystyle{IEEEtran}
\bibliography{bib}

\end{document}